\documentclass[aps, prd, twocolumn, groupedaddress,showpacs, nofootinbib]{revtex4-1}

\usepackage[english]{babel}
\usepackage[T1]{fontenc}
\usepackage[utf8]{inputenc}

\usepackage{amsmath}
\usepackage{graphicx}
\usepackage{amssymb}
\usepackage{multirow}
\usepackage{subfigure}
\usepackage{xspace}
\usepackage{enumerate}
\usepackage[dvipsnames]{xcolor}

\newcommand{\lsca}{\lambda_{\mathrm{scatt}}}
\newcommand{\lcoh}{\lambda_{\mathrm{coh}}}
\newcommand{\Rmax}{R_{\mathrm{max}}}
\newcommand{\Dmin}{D_{\mathrm{min}}}
\newcommand{\Efit}{E_{\mathrm{min}}^{\mathrm{fit}}}
\newcommand{\desy}{{DESY, D-15738 Zeuthen, Germany}}

\newcommand{\ifsc}{{Instituto de Física de São Carlos, Universidade de São Paulo, \\ Av. Trabalhador São-Carlense, 400, São Carlos, SP, Brazil}}
\newcommand{\nbi}{{Niels Bohr International Academy \& Discovery Center, Niels Bohr Institute,\\University of Copenhagen, Blegdamsvej 17, DK-2100 Copenhagen, Denmark}}

\begin{document}

\title{Revisiting the distance to the nearest UHECR source:\\ Effects of extra-galactic magnetic fields}

\author{Rodrigo~Guedes~Lang}
\email{rodrigo.lang@usp.br}
\affiliation{\ifsc}

\author{Andrew~M.~Taylor}
\affiliation{\desy}

\author{Markus~Ahlers}
\affiliation{\nbi}

\author{Vitor~de~Souza}
\affiliation{\ifsc}

\date{\today}

\begin{abstract}
We update the constraints on the location of the nearest UHECR source. By analyzing recent data from the Pierre Auger Observatory using state-of-the-art CR propagation models, we reaffirm the need of local sources with a distance less than 25-100~Mpc, depending on mass composition. A new fast semi-analytical method for the propagation of UHECR in environments with turbulent magnetic fields is developed. The onset of an enhancement and a low-energy magnetic horizon of cosmic rays from sources located within a particular distance range is demonstrated. We investigate the distance to the nearest source, taking into account these magnetic field effects. The results obtained highlight the robustness of our constrained distances to the nearest source.
\end{abstract}

\maketitle

\section{Introduction}
\label{sec:introduction}

The origin of ultra-high energy cosmic rays (UHECR) remains an open question even a century after their discovery~\cite{AlvesBatista:2019tlv}. Deflections of cosmic rays in extra-galactic and galactic magnetic fields scramble their arrival direction and, consequently, mask the location of their sources. Only for the most energetic events ($E \gtrsim 10^{19.5}$~eV), some residual information about their origin can still be present in their arrival direction distribution. In particular, a recent analysis of data from the Pierre Auger Observatory~\cite{PierreAuger} has revealed a strong large-scale dipole anisotropy at the level of 6.5\% above 8~EeV~\cite{AugerDipole} (see also~\cite{Ahlers:2018qsm,Aab:2020xgf}). In addition, the data above 39~EeV shows hints of a cross-correlation with $\gamma$-ray data, in particular, for the sub-sample of starburst galaxies~\cite{AugerStarburst}. While these results can be considered important milestones towards the identification of UHECR sources, the overall data is presently inconclusive. We refer to the recent reviews~\cite{Ahlers:2016rox,Deligny:2018blo} for further details.

The energy spectrum of UHECR, on the other hand, has been measured with unprecedented statistics~\cite{AugerSpectrum}, revealing two important features: a hardening of the spectral index at $E = 10^{18.8}$~eV, the so-called ankle, and a suppression for $E > 10^{19.7}$~eV~\cite{AugerSpectrumICRC}, which may be explained by either a maximum power of acceleration of the sources or energy losses during the propagation or a combination of both~\cite{AugerCombinedFit}. During their propagation, UHECR interact with the photon background and lose energy via $e^-e^+$ pair production, pion production and photodisintegration, the latter of which leads to a change in the particle species~\cite{Hooper:2006tn,Allard:2011aa}. These losses are energy dependent, with the highest energies ($E \gtrsim 10^{19.5}$~eV) being dominated by pion production for a proton and by photodisintegration for heavier nuclei, the so-called GZK effect~\cite{PhysRevLett.16.748,Zatsepin:1966jv}. These propagation effects create an energy-dependent horizon, dictating a maximum distance from which UHECR of a given energy are expected to come from~\cite{Lemoine_2005,Aloisio_2005,Globus_2007,Mollerach_2013}. As a consequence, the energy spectrum also contains information about the distance distribution of the sources of UHECR, which might be helpful in deciphering their origins.

In this work, we investigate the role played by the local sources of UHECR on the observed spectrum. Firstly, in section~\ref{sec:Dmin}, we revisit the work of Ref.~\cite{Taylor11}, that constrained the distance to the nearest source for an environment with no magnetic fields. This study is updated here by analyzing recent data from the Pierre Auger Observatory using the state-of-the-art Monte Carlo propagation code, CRPropa 3~\cite{CRPropa3}. We also discuss the stability and systematic uncertainties of the fit. We develop the analysis by discussing the combined effects of a distance to the nearest source and extra-galactic turbulent magnetic fields. In section~\ref{sec:method}, we describe a semi-analytical approximation for the propagation of UHECR in turbulent extra-galactic magnetic fields, that provides an efficient method of studying the effect of a magnetic horizon in the cosmic ray data analysis. In section~\ref{sec:horizon}, we explore the effects of the distance to the nearest source in the low-energy end of the spectrum for different magnetic field scenarios. The maximum distance to the nearest source is again constrained for such scenarios. Finally, we conclude in section~\ref{sec:conclusions}.

\section{Distance to the nearest source}
\label{sec:Dmin}

Nearby UHECR sources are necessary to explain the high-energy end of the cosmic ray spectrum ($E
\gtrsim 10^{19.5}$~eV) due to the energy-loss horizon of these particles. A better understanding of this requirement may provide additional information about the UHECR source distribution. We update the work of Ref.~\cite{Taylor11} by analyzing recent UHECR data from the Pierre Auger Observatory. 

We consider spatially uniform distributions of sources that accelerate UHECR at a constant rate. The injection spectrum follows a power law with spectral index $\Gamma$ and a rigidity-dependent exponential cutoff, $\exp(-R/\Rmax)$. We assume a pure mass composition at the sources, with five representative primaries, $^1$H, $^{4}$He, $^{14}$N, $^{28}$Si and $^{56}$Fe. Such a species range has an approximately uniform spacing in $\ln A$, where $A$ is the atomic mass. We start as in Ref.~\cite{Taylor11} by neglecting the effects of magnetic fields and consider the robustness of the results under more general conditions in the following section. In the absence of magnetic fields, the use a one-dimensional (1-D) UHECR propagation treatment is justified.

\begin{figure}[ht]
    \centering
    \includegraphics[width=0.49\textwidth]{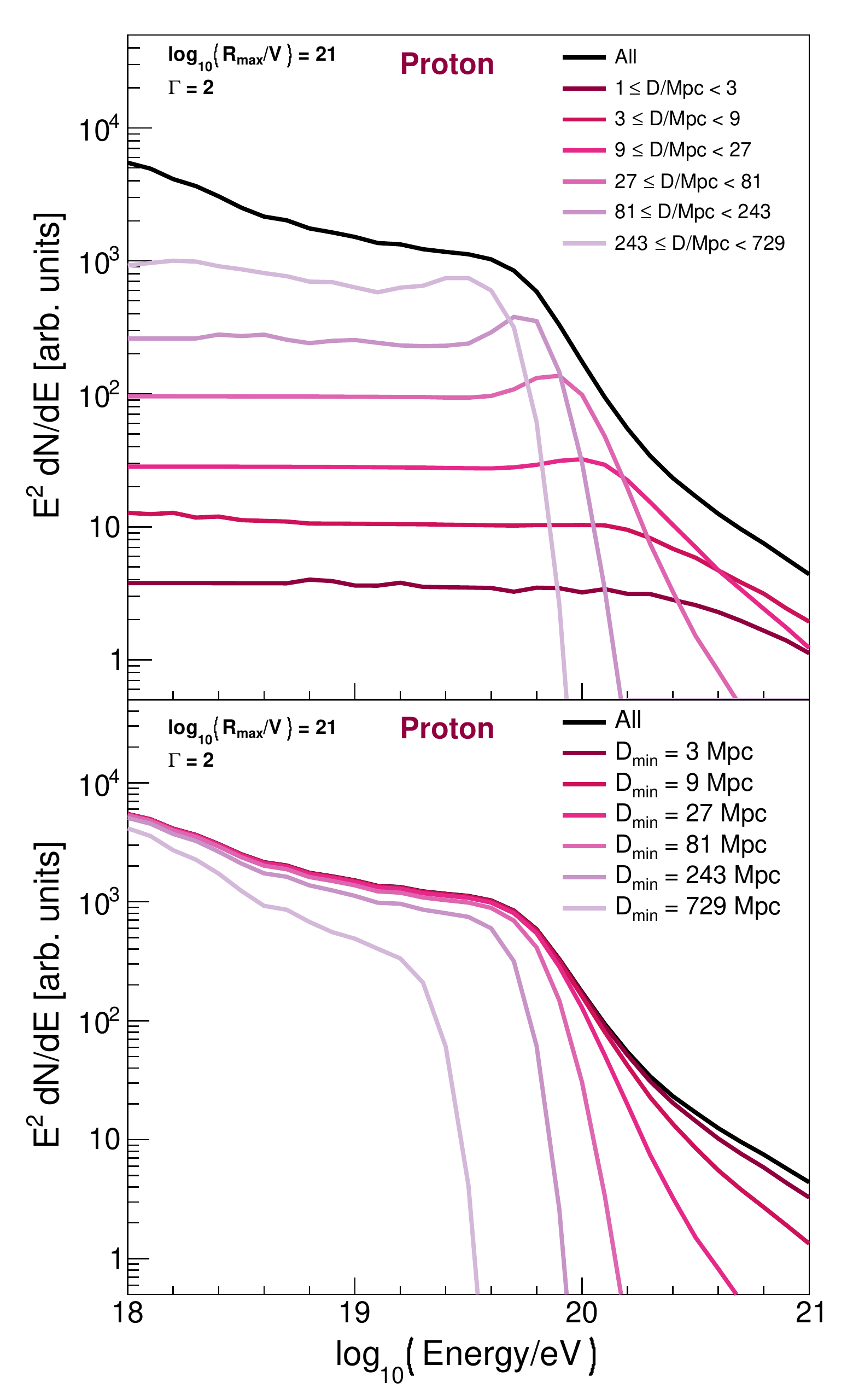}
    \caption{Example spectrum of cosmic rays divided into distance shells with protons as primaries. On the top panel, the black line represents the total flux, while the colored lines represent the contribution of each distance shell. On the bottom panel, on the other hand, each colored line shows the total flux for a given distance to the nearest source, $\Dmin$.}
    \label{fig:ShellsExample_p}
\end{figure}

\begin{figure}[!h]
    \centering
    \includegraphics[width=0.49\textwidth]{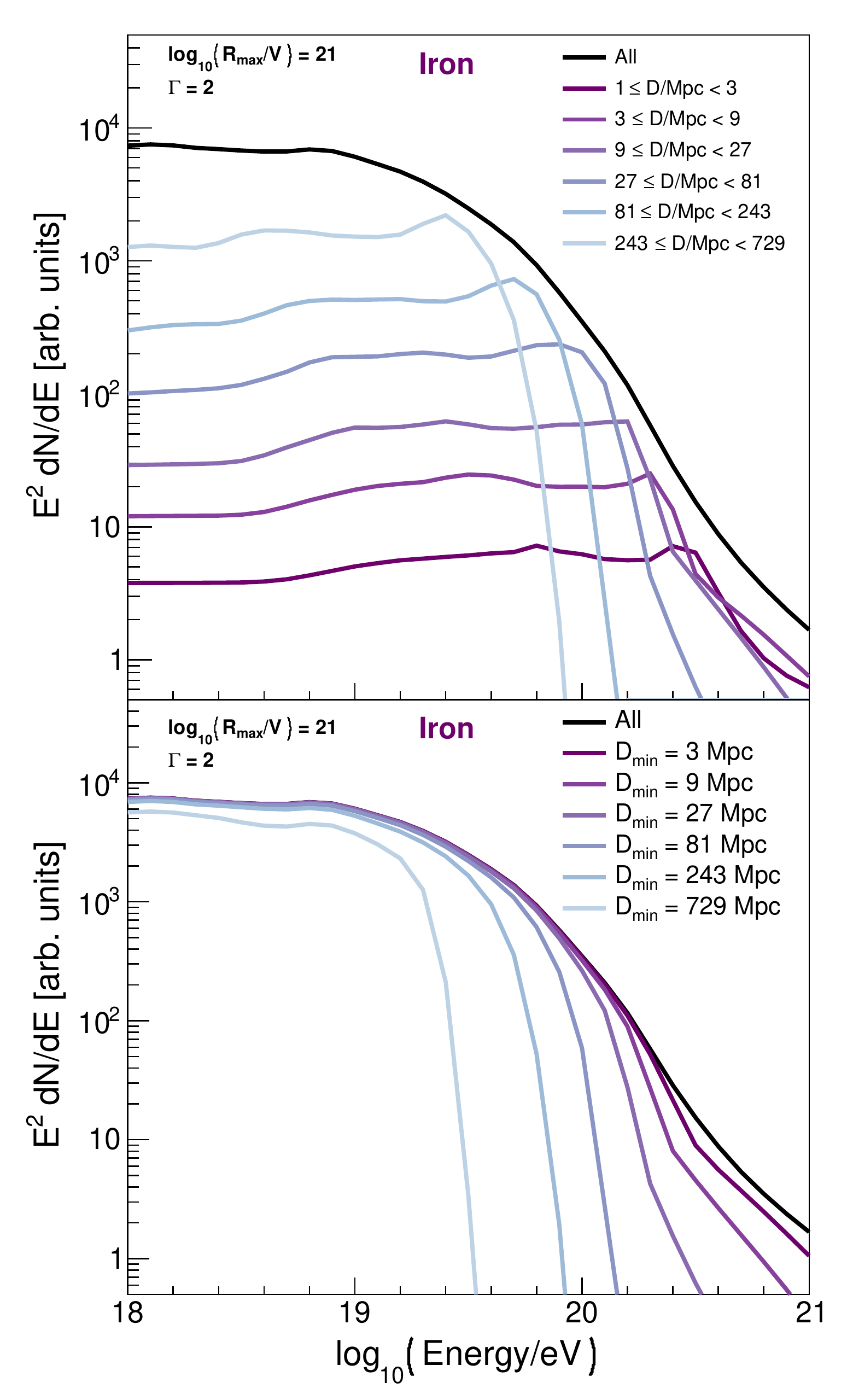}
    \caption{Same as figure~\ref{fig:ShellsExample_p}, but for iron.}
    \label{fig:ShellsExample_Fe}
\end{figure}

Using the state-of-the-art Monte Carlo propagation code CRPropa 3, we obtain the arriving spectrum from an ensemble of sources. A simple example scenario with $\Gamma = 2$ and $\Rmax = 10^{21}$~V is considered in order to qualitatively highlight these effects. Figures~\ref{fig:ShellsExample_p} and \ref{fig:ShellsExample_Fe} show the spectra originating from different distance shells as well as the resulting spectrum for a given distance to the nearest source, $\Dmin$, for the two extreme primaries, proton and iron. Each distance shell dominates a different energy range in such a way that local sources contribute the most to the very end of the spectrum and, thus, large values of $\Dmin$ lead to a strong suppression of the flux, which is not compatible with the data.

\subsection{Analysis method}
\label{subsec:fit}

To quantify the proximity of the most local UHECR sources, we fit for each primary the spectral data of the Pierre Auger Observatory~\cite{AugerSpectrumICRC} using a simple $\chi^2$-test. We test different values of the minimum energy bin of the fit, $\Efit$. For each value of $\Efit$ and $\Dmin$, the spectral parameters, $\Gamma$ and $\Rmax$, as well as the normalization are taken as free parameters and fitted to the data. We account for the systematic uncertainties by perfoming a scan in the energy scale from -14\% to 14\%. The effects of $\Dmin$ due to propagation losses reveal themselves at the highest energies, at which the particles propagate almost ballistically in the magnetic fields.

The distance to the nearest source has much stronger effects on the measured spectrum than on the measured composition. Similar results are found for every primary. Therefore, a simple fit of the spectral data for a pure composition scenario is sufficient to address the main effects.

\subsection{Maximal distances of the nearest source}

\begin{figure}[!h]
    \centering
    \includegraphics[width=0.49\textwidth]{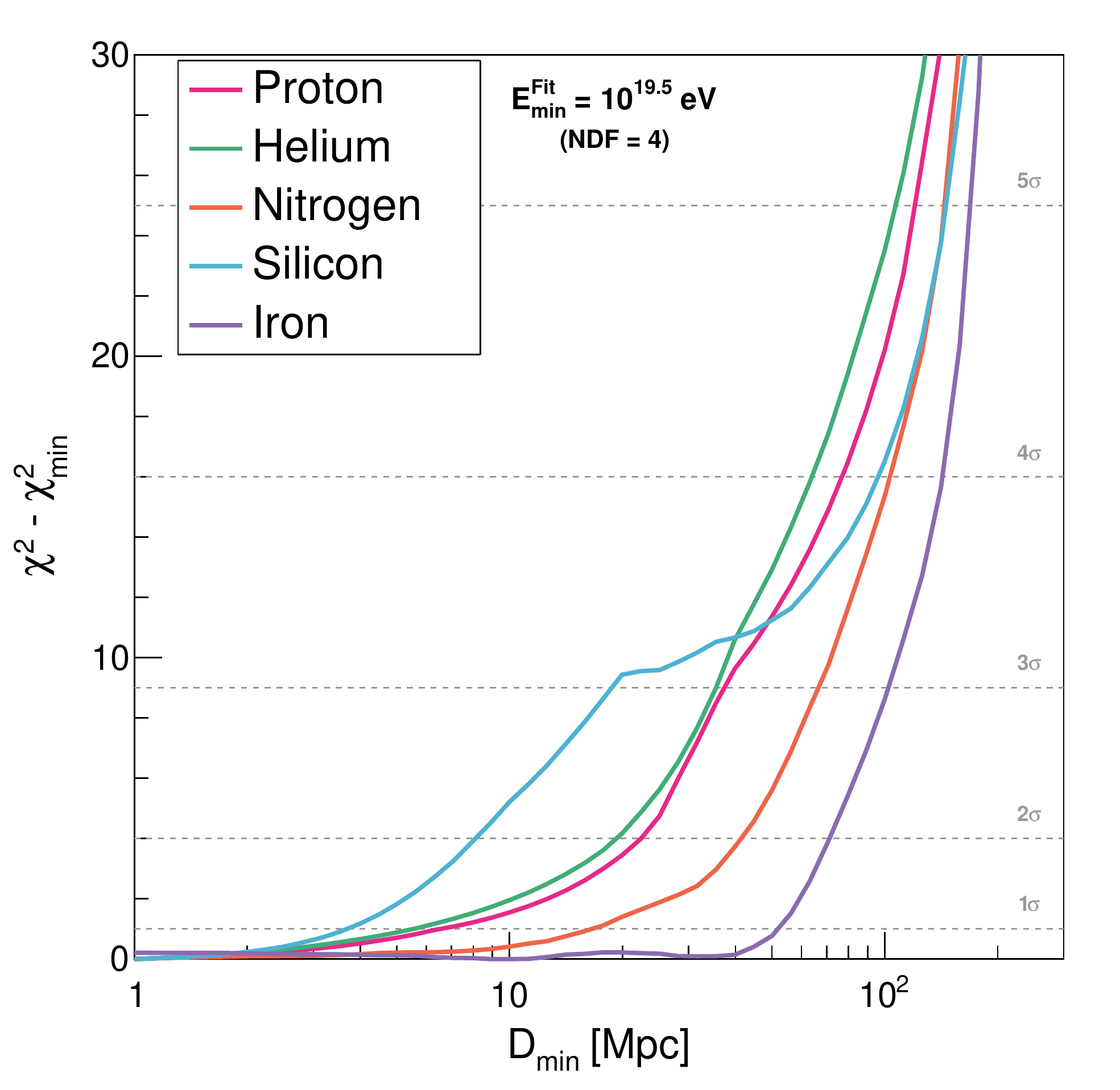}
    \caption{Value of $\chi^2$ obtained for the best fit parameters for each distance to the nearest source, $\Dmin$. The results for $\Efit = 10^{19.5}$~eV are shown, similar distributions were obtained for other values of $\Efit$. The pink, green, orange, cyan and purple lines represent, respectively, the scenarios with pure proton, helium, nitrogen, silicon and iron composition at the sources. The confidence level of rejection, $\sigma$, is shown for comparison.}
    \label{fig:chi2}
\end{figure}

\begin{figure}[!h]
    \centering
    \includegraphics[width=0.49\textwidth]{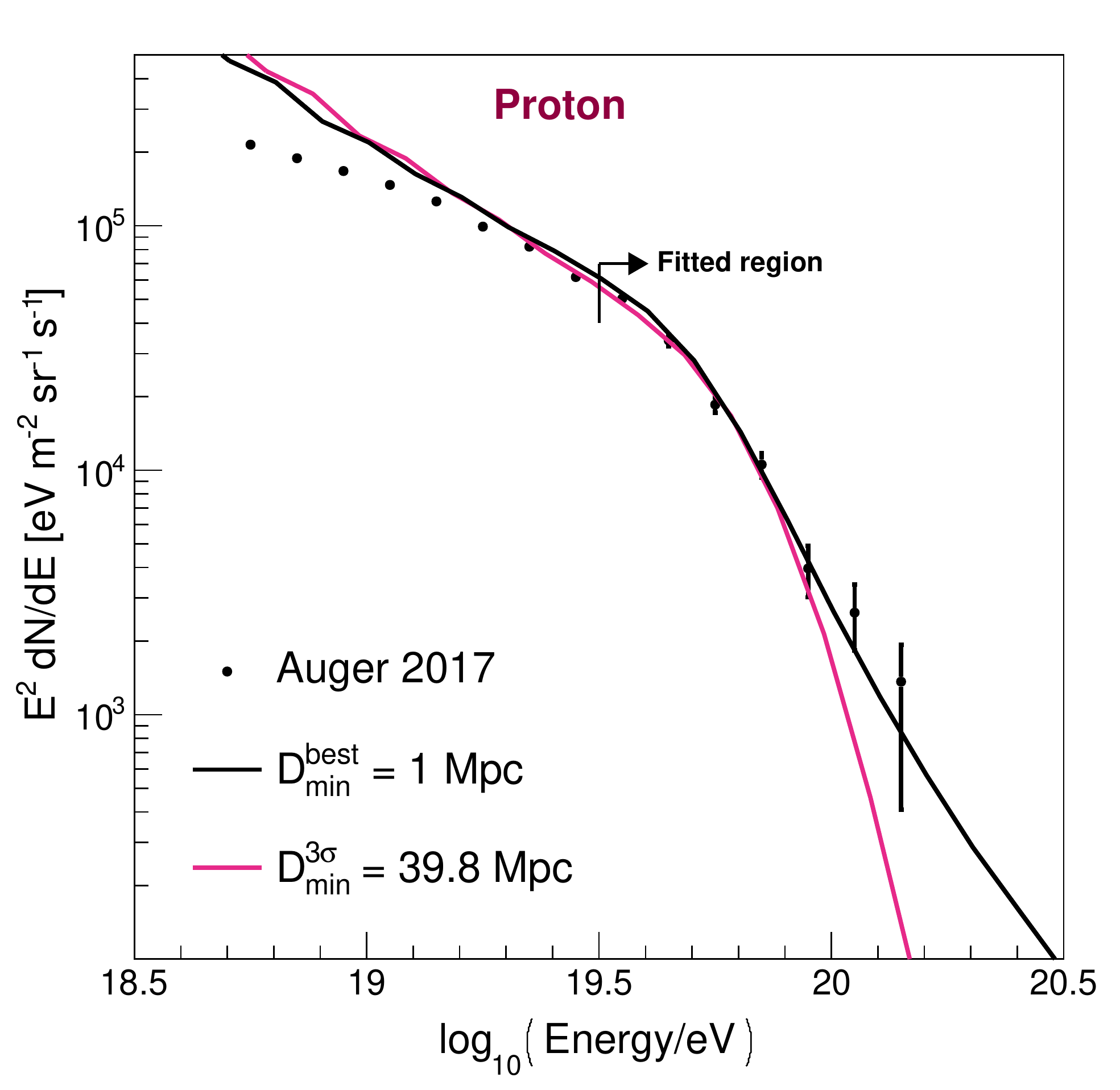}
    \includegraphics[width=0.49\textwidth]{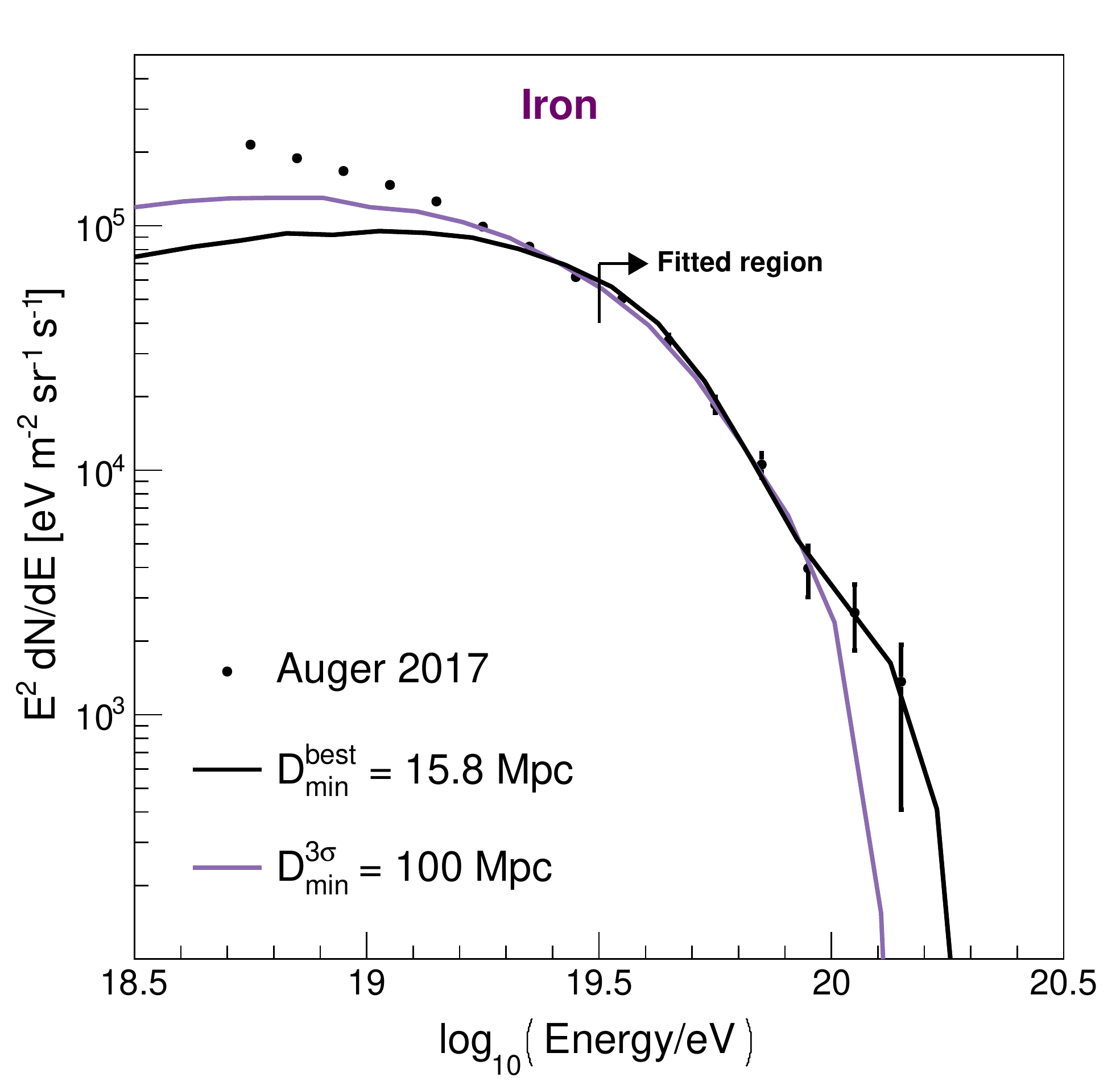}
    \caption{Spectrum of the best fit scenario for both $\Dmin^{\mathrm{best}}$ and $\Dmin^{3\sigma}$. The top and bottom panel are for pure proton and iron composition respectively. The black lines show the best-fit spectra for the values of $\Dmin$ which best describes the data, while the colored lines show the best-fit spectra for $\Dmin^{3\sigma}$. The spectral parameters for $\Dmin^{3\sigma}$ are shown in Table
   ~\ref{tab:dmin}, while the spectral parameters for $\Dmin^{\mathrm{best}}$ are ($\Gamma^{\mathrm{best}} = 2.8$, $\log_{10} (\Rmax^{\mathrm{best}}/\mathrm{V}) = 20.5$) and ($\Gamma^{\mathrm{best}} = 0.8$, $\log_{10} (\Rmax^{\mathrm{best}}/\mathrm{V}) = 19.7$) for proton and iron respectively.}
    \label{fig:spectrumfit}
\end{figure}

Figure~\ref{fig:chi2} shows the evolution of $\Delta \chi^2$ as the value of $\Dmin$ is increased for $\Efit = 10^{19.5}$~eV. The data is best described by small values of $\Dmin$, reinforcing the need for local sources. Large distances to the nearest source can be statistically rejected with a confidence level given by $\sigma = \sqrt{\chi^2 - \chi^2_{\mathrm{min}}}$. The reference rejected distances at 3$\sigma$ (99.7\%) confidence level for $\Efit = 10^{19.5}$~eV are shown in Table~\ref{tab:dmin}. The sudden change in the behavior of the $\chi^2$ distribution for silicon comes from the combination of the dependency of the photodisintegration cross section and energy threshold with the mass and the energy in which the corresponding $\Dmin$ shell is dominant.

Figure~\ref{fig:spectrumfit} shows the resulting spectra for the best fit scenario, $\Dmin^{\mathrm{best}}$, as well as the scenario rejected at 3$\sigma$ confidence level, $\Dmin^{3\sigma}$. Large distances to the nearest source result in a severe suppression at the highest energies due to the UHECR horizon, which is in disagreement with experimental data. Similar effects on the spectrum were found for helium, nitrogen and silicon.

\begin{table}[ht]
    \centering
    \begin{tabular}{c|c|c|c}
    \hline
    \hline
    Primary & $\Dmin^{3\sigma}$ [Mpc] & $\Gamma^{3\sigma}$ & $\log_{10}(\Rmax^{3\sigma} / \mathrm{V})$\\
    \hline
        p & 40 & 2.8 & 20 \\
        He & 40 & 2.6 & 23 \\
        N & 70 & 2.7 & 23 \\
        Si & 31 & 2.6 & 23 \\
        Fe & 100 & 0.8 & 19.5 \\
        \hline
        \hline
    \end{tabular}
    \caption{Reference rejected distances to the nearest source of UHECR at 3$\sigma$ (99.7\%) confidence level and the fitted spectral parameters for such cases. A minimum energy of the fit of $\Efit = 10^{19.5}$~eV is considered.}
    \label{tab:dmin}
\end{table}

\subsection{Systematics}

\begin{figure}[!h]
    \centering
    \includegraphics[width=0.49\textwidth]{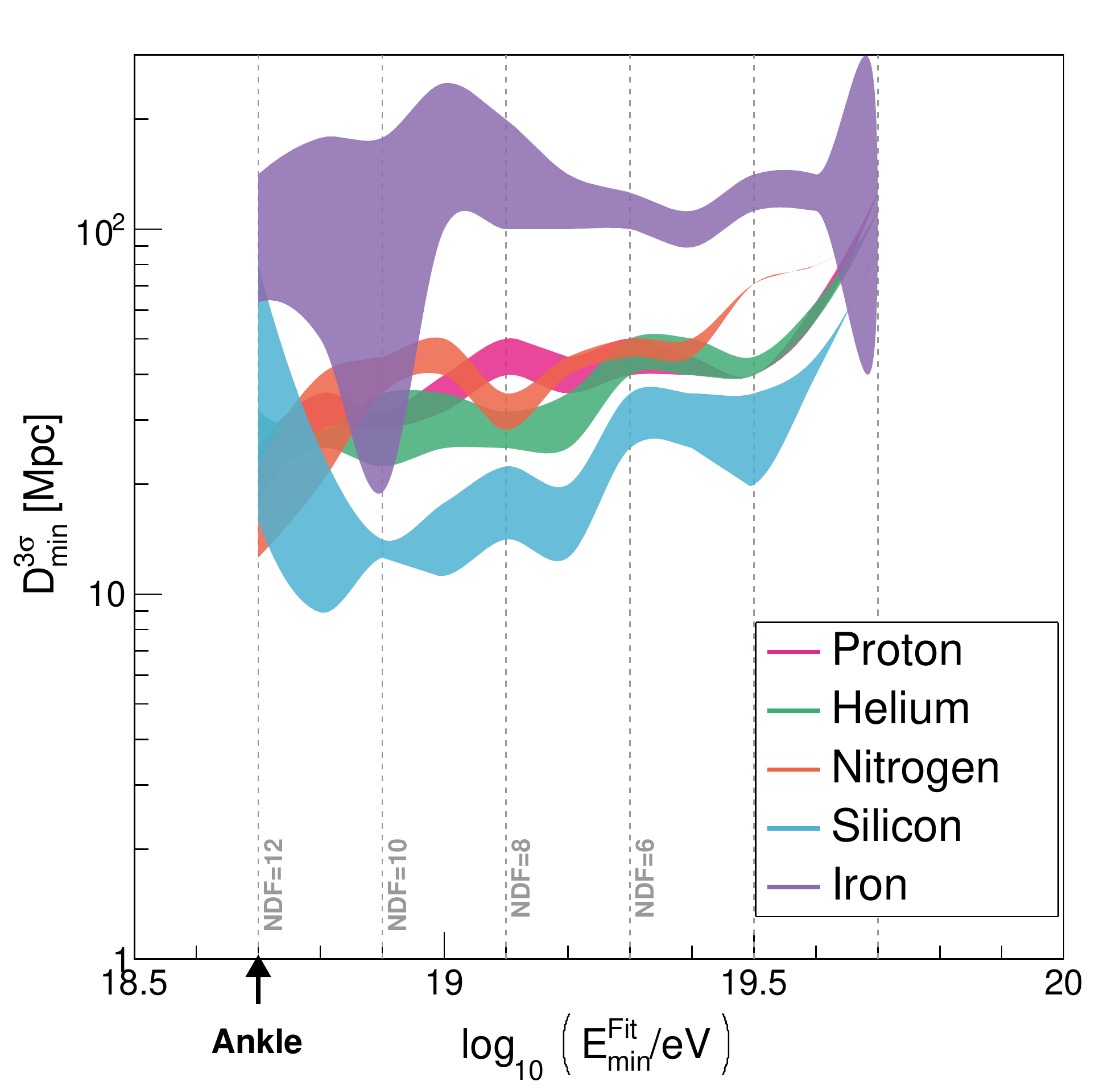}
    \caption{Upper limits of the distance to the nearest source at 3$\sigma$ (99.7\%) for increasing value of $\Efit$. The pink, green, orange, cyan and purple lines represents, respectively, the scenarios with pure proton, helium, nitrogen, silicon and iron primaries. The number of degrees of freedom (NDF) for each $\Efit$ are shown by the dashed gray lines. The area represents the uncertainty coming for the EBL model. The Kneiske model~\cite{Kneiske_2004} as well as the upper and lower limits on the distribution coming from the Stecker model~\cite{Stecker_2016} were considered.}
    \label{fig:chi2fitmin}
\end{figure}

We investigate the systematic uncertainty of the analysis by evaluating the influence of some of the model assumptions in the final result, i.e., $\Dmin^{3\sigma}$. In particular we address the minimum energy considered in the fit, $\Efit$, the source evolution, and the extra-galactic background light model (EBL).

Cosmic rays interact with the background radiation fields, including the EBL, resulting in energy losses. The EBL distribution, however, is not well understood and several competitive models are used to describe it. In this work, we use the Kneiske model~\cite{Kneiske_2004} and the upper and lower limits from the Stecker model~\cite{Stecker_2016}, as representative EBL distributions.

The source evolution is modeled as $(1+z)^k$ and the systematics coming from it are insignificant. Even for a strong source evolution with redshift such as those following star formation rates, in which the number of sources grow with $(1+z)^{3.6}$ for small values of redshift~\cite{Hopkins:2006bw,Y_ksel_2008}, the results are exactly the same. This is expected, since the studied effects are coming from close sources ($z \lesssim 0.02$), for which the density of sources would change only by a factor of $1.02^{3.6} \approx 1.07$.

Figure~\ref{fig:chi2fitmin} shows how the value of $\Dmin^{3\sigma}$ changes for different values of $\Efit$ and EBL models for each primary. While the resulting fit parameters depend heavily on the assumptions of the fit~\cite{AugerCombinedFit}, the inferred value for $\Dmin^{3\sigma}$ proves to be stable in relation to these parameters, and consequently even a simple fit such as the one proposed here can be used to obtain important insights on the local sources of UHECR. Considerations of more realistic mixed composition would lead to similar conclusions with restrained distances to the nearest source lying somewhere in between the results obtained in this analysis.

\section{Magnetic fields}

We further study the constraints on the distance to the nearest source by investigating the effects of the presence of turbulent extra-galactic magnetic fields.

Being charged particles, UHECR are deflected by both galactic and extra-galactic magnetic fields. Although this can impact the distribution of arrival directions, it is expected that the total flux from a homogeneous distribution of sources will not be changed by these deflections, as it has previously been deduced from the application of Liouville's theorem~\cite{Aloisio_2004}.

Nevertheless, the relative contribution to the total cosmic ray spectrum from sources in each distance shell is strongly dependent on the extra-galactic magnetic field strength. The distance to the nearest source can thus have an impact on both the total arriving spectrum and composition at different energies, for scenarios in which strong extra-galactic magnetic fields exist.

In order to further investigate this, we first present a semi-analytical method for obtaining the UHECR spectrum and the corresponding contribution of each distance shell in such environments. Subsequently, we discuss the resulting effects and finally study the robustness of the maximum distance to the nearest sources in the presence of extra-galactic magnetic fields (EGMF).

\subsection{Semi-analytical propagation method}
\label{sec:method}

The propagation of ultra-high energy nuclei is stochastic in nature and, thus, can be studied with Monte Carlo methods. 
Environments without magnetic fields can be efficiently simulated by adopting simplifying techniques such as performing the simulation in 1-D and reweighting the events using the sources energy and distance distributions. Nevertheless, when a general extragalactic magnetic field is considered, a so-called 4-D simulation is needed, taking into account the spatial scales and also the time (or redshift) at which the cosmic ray was emitted. This increases the computational cost, due to the extra dimensions considered as well as to the fact that most of the simulated cosmic rays do not arrive at Earth.

If the considered fields are turbulent and isotropic, however, the propagation remains radially symmetric around each source and, thus, a mapping of the 1-D Monte Carlo simulation into a 4-D result is possible.

In order to do so, it is necessary to obtain the distance distribution of cosmic rays from each source as a function of time, $dN/dr (t, \lsca)$, where $\lsca (R,B,\lcoh)$ is the scattering length.

In this work, we use a simple prescription for the scattering length, which is motivated by present limitations in our knowledge of the actual field structures,

\begin{equation}
    \label{eq:lint}
    \lsca = \begin{cases}
    \left(\frac{R_{L}}{\lcoh}\right)^{1/3} \lcoh &\mathrm{for} \ R_{L} < \lcoh\, \\
    \left(\frac{R_{L}}{\lcoh}\right)^{2} \lcoh, &\mathrm{for} \ R_{L} \ge \lcoh\,
    \end{cases},
\end{equation}
where $\lcoh$ is the coherence length of the field and $R_{L}$ is the Larmor radius of the particle given by
\begin{equation}
    R_{L} = \frac{p}{|q|B} \approx \frac{1.081}{Z} \left(\frac{E}{\mathrm{EeV}}\right) \left(\frac{\mathrm{nG}}{B}\right) \ \mathrm{Mpc}.
\end{equation}

Three regimes are considered depending on the rigidity, travel time and magnetic field properties: for short times the cosmic ray propagates ballistically and a simple delta function is enough to describe the distribution; for large times, the propagation is diffusive and a truncated Gaussian is used, and for the intermediate regime a J\"uttner distribution is needed. 

The 1-D Monte Carlo simulations are then mapped into a 4-D result by an analytic expression for the fraction $P$ of cosmic rays emitted at time $t$ (or equivalently distance $D=ct$ in the 1-D simulation) that were emitted at sources in a distance window $(D_{\mathrm{min}}, D_{\mathrm{max}})$, which is given by:

\begin{equation}
    P = \int_{D_{\mathrm{min}}}^{D_{\mathrm{max}}} \frac{dN}{dr} (t, \lsca) dr.
\end{equation}
The distribution $dN/dr$ is explained in details in Appendix~\ref{app:distributions} and an example distribution of this function is shown in figure~\ref{fig:DistanceDistribution}. Additionally, the simulation setup, the mapping, and the validation of the method are described in Appendix~\ref{app:setup}. The method obtains results consistent to those obtained by a full 4-D simulation, with more than 4 orders of magnitude less computational time and a better control of the simulation parameters.

For the mapping of propagation time to distance, we assume that the scattering length of the cosmic rays remains approximately constant during their propagation. This is reasonable, since the main energy loss mechanism for nuclei is photodisintegration, in which the rigidity may change by a factor of $26/56 \approx 0.5$ in the worst case. For protons at the lowest energies considered, the main energy loss mechanism is the $e^-e^+$ pair production which has a large loss length and has previously been demonstrated to be safely neglected~\cite{Mollerach_2013}. 

\subsection{Spectral effects of extra-galactic magnetic fields}
\label{sec:horizon}

In this section, we decipher the various spectral features expected to arise in an environment with magnetic fields, following the procedure described in section~\ref{sec:method}.

We consider an extra-galactic magnetic field with Kolmogorov turbulence power-spectrum (see Appendix~\ref{app:setup} for more details). Such a field construction can be fully characterized by its RMS field strength, $B$, and coherence length, $\lcoh$. We illustrate the effects with a representative scenario using $\Gamma = 2$, $\Rmax = 10^{21}$~V and $B = 3$~nG. This is an example scenario that qualitatively highlights the effects, more realistic scenarios are considered further on the analysis.

\begin{figure}[ht!]
    \centering
    \includegraphics[width=0.49\textwidth]{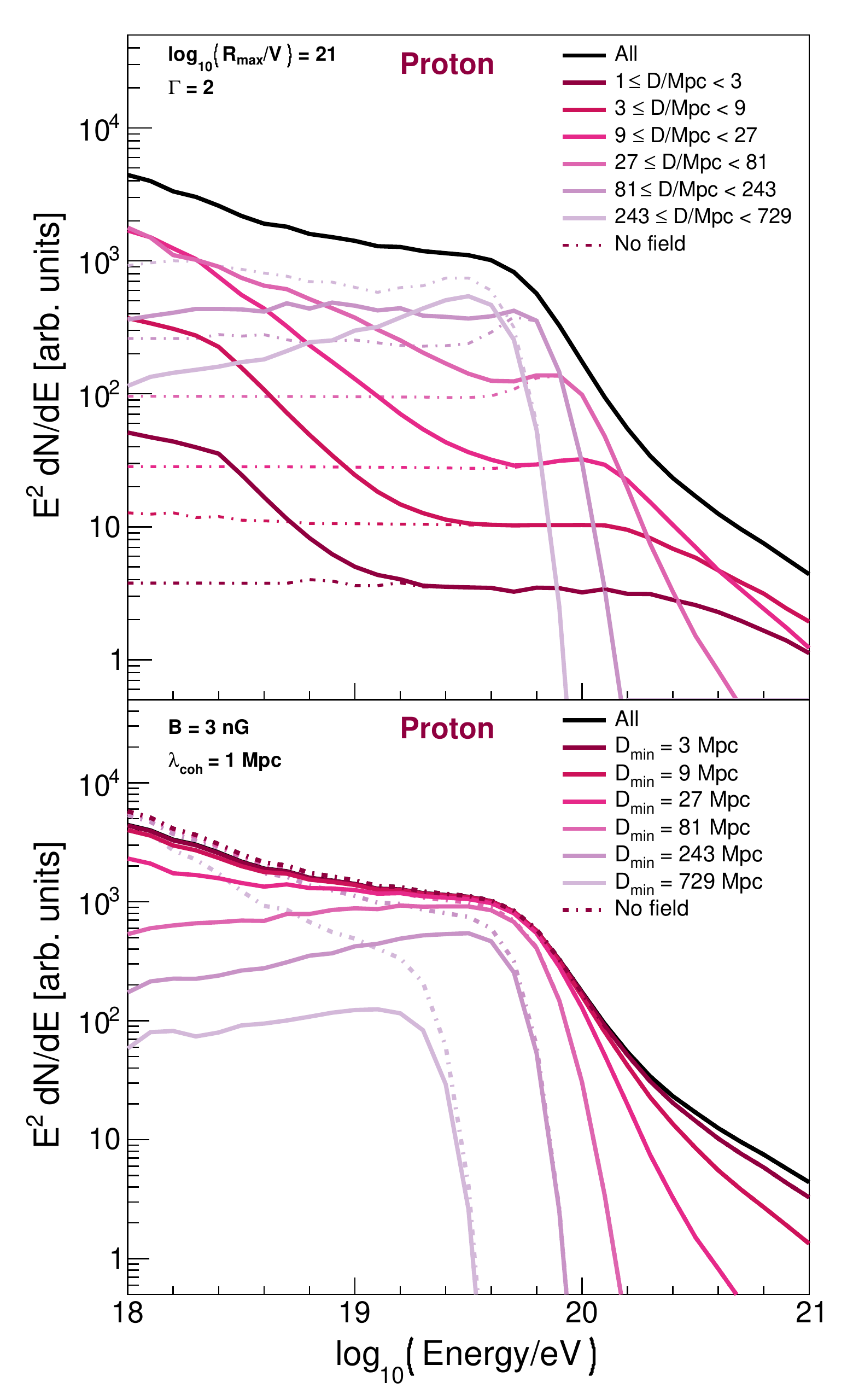}
    \caption{Spectrum of cosmic rays in an environment with turbulent magnetic fields divided into distance shells for a pure composition of proton at the sources. The parameters of the fields are taken as $B = 3$~nG and $\lcoh = 1$~Mpc. On the top panel, each colored line shows the contribution from each distance shell, while the black line shows the total spectrum. On the bottom panel, each colored line represent a scenario with a different distance to the nearest source, $\Dmin$. The dashed lines show the case with no magnetic fields for comparison.}
    \label{fig:ShellsExampleMagneticFields_p}
\end{figure}

\begin{figure}[ht!]
    \centering
    \includegraphics[width=0.49\textwidth]{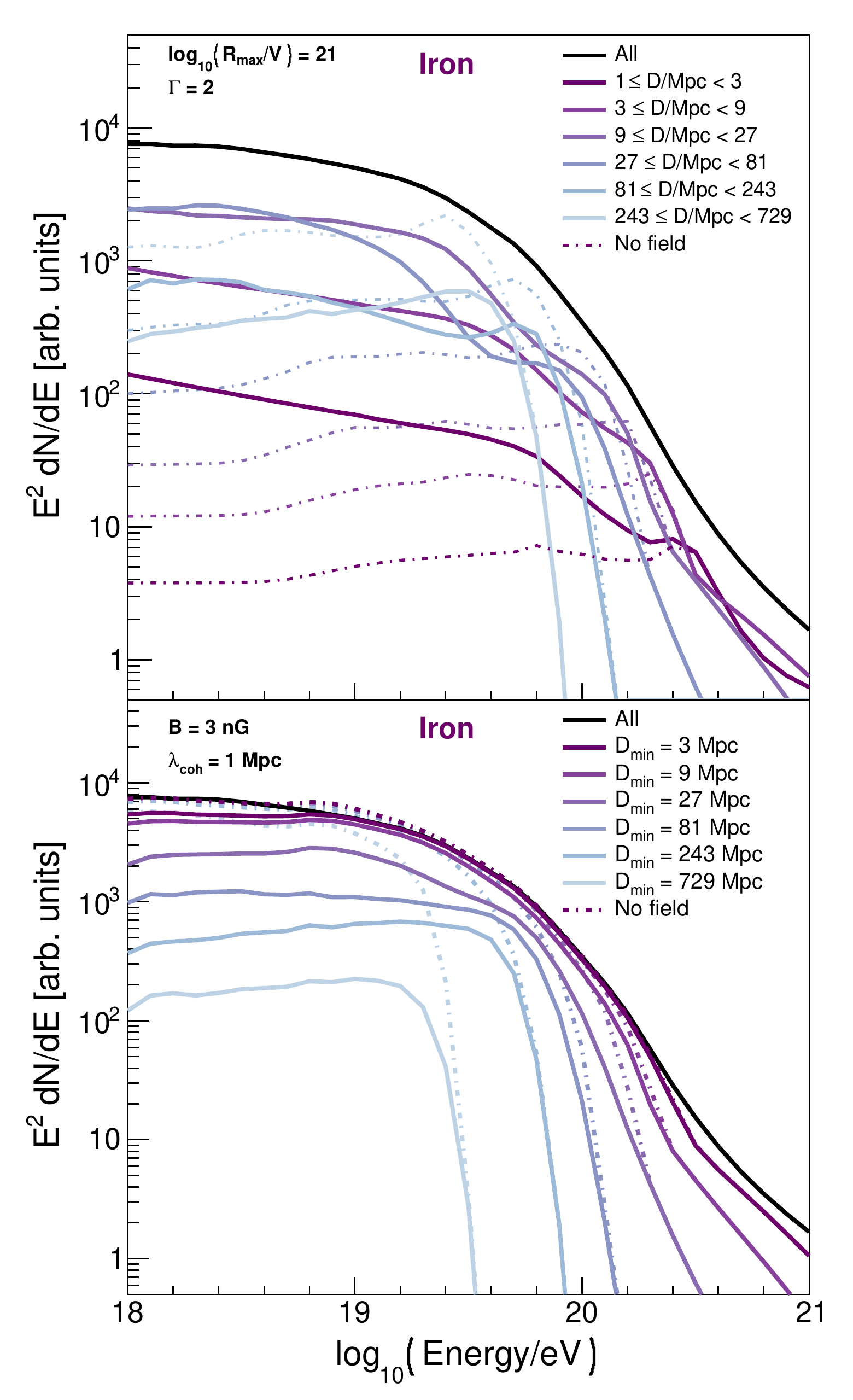}
    \caption{Same as figure~\ref{fig:ShellsExampleMagneticFields_p}, but for iron.}
    \label{fig:ShellsExampleMagneticFields_Fe}
\end{figure}

The effects of magnetic fields on the contribution of each distance shell can be seen in the upper panels of figures~\ref{fig:ShellsExampleMagneticFields_p} and \ref{fig:ShellsExampleMagneticFields_Fe}, for proton and iron, respectively. As visible in the individual plots, up to four regimes are present in each shell and are convoluted with the spectral features coming from energy loss processes. From higher to lower energies, they can be defined as:

\begin{enumerate}[I]
    \item \textbf{Ballistic:} at the highest energies, the rigidity is sufficiently high, and the propagation distances sufficiently short, such that magnetic field effects do not arise.
    \item \textbf{Non-resonant scattering enhancement:} for lower energies, the cosmic ray number density from a shell increases due to the accumulation of particles scattering over time and their ability to return to a region. In this region, the enhancement scales as as $E^{-2}$.
    \item \textbf{Resonant scattering enhancement:} due to the energy dependence of the scattering lengths, which is encapsulated in equation~\ref{eq:lint}, in this region the enhancement scales as $E^{-1/3}$.
    \item \textbf{Magnetic horizon:} a low-energy magnetic horizon emerges due to the finite age of the Universe.
\end{enumerate}

The farther the shell, the higher the energy up to which these effects are manifest.

In the lower panels of figures~\ref{fig:ShellsExampleMagneticFields_p} and \ref{fig:ShellsExampleMagneticFields_Fe}, the total spectrum for a given distance to the nearest source is shown. The combination of the four effects aforementioned in each shell results in a notable low-energy horizon and a hardening of the spectrum above the ankle and close to GZK energies. For some combinations of field intensity and distance to the nearest source, a significant change in the arriving composition is expected. Such a change may potentially account for the large fraction of protons observed below the ankle~\cite{AugerXmax}, though the investigation of such a possibility lies beyond the scope of this study.

\subsection{Distance constraints with magnetic fields}

\begin{figure}[h!]
    \centering
    \includegraphics[width=0.45\textwidth]{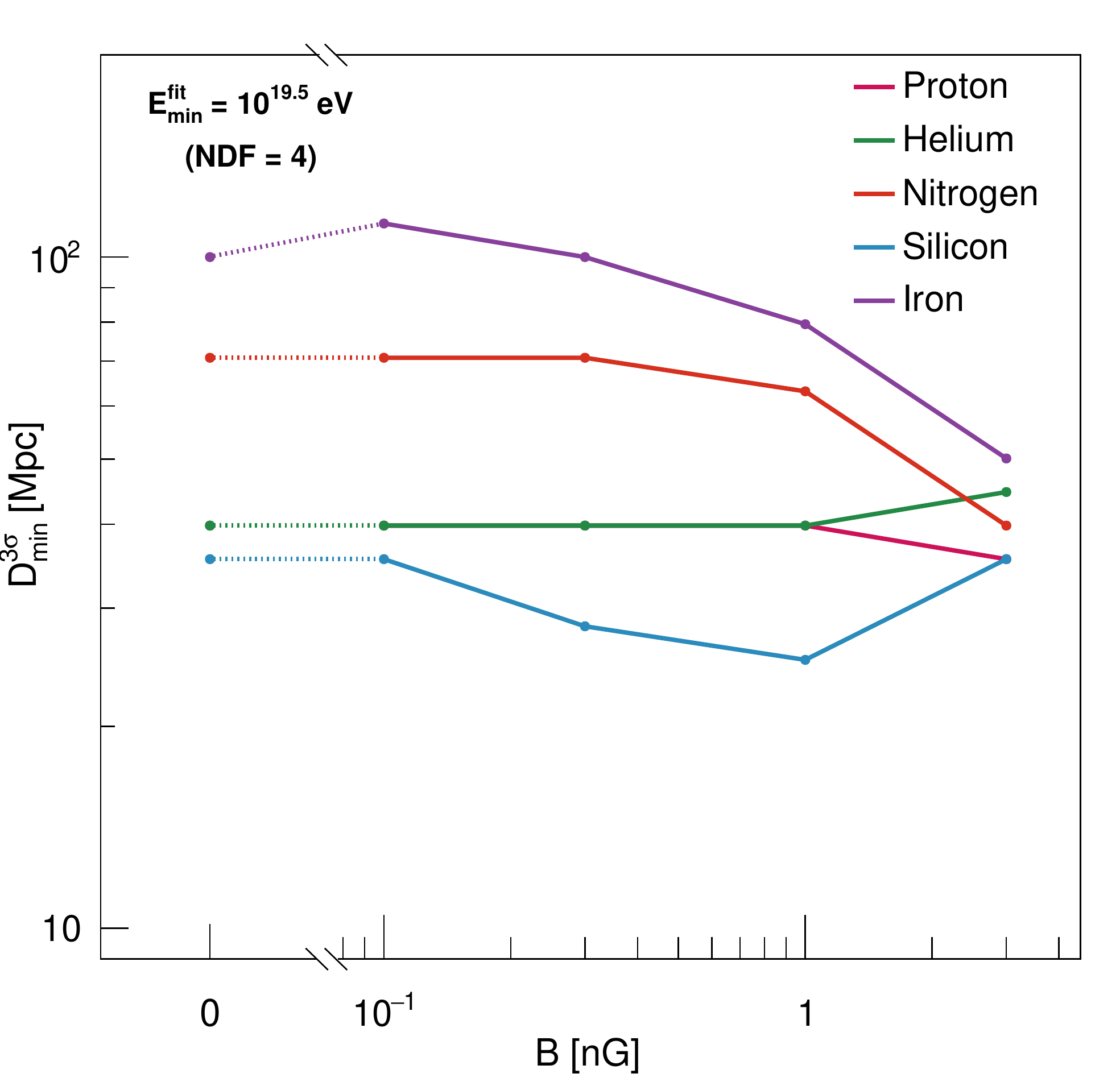}
    \includegraphics[width=0.45\textwidth]{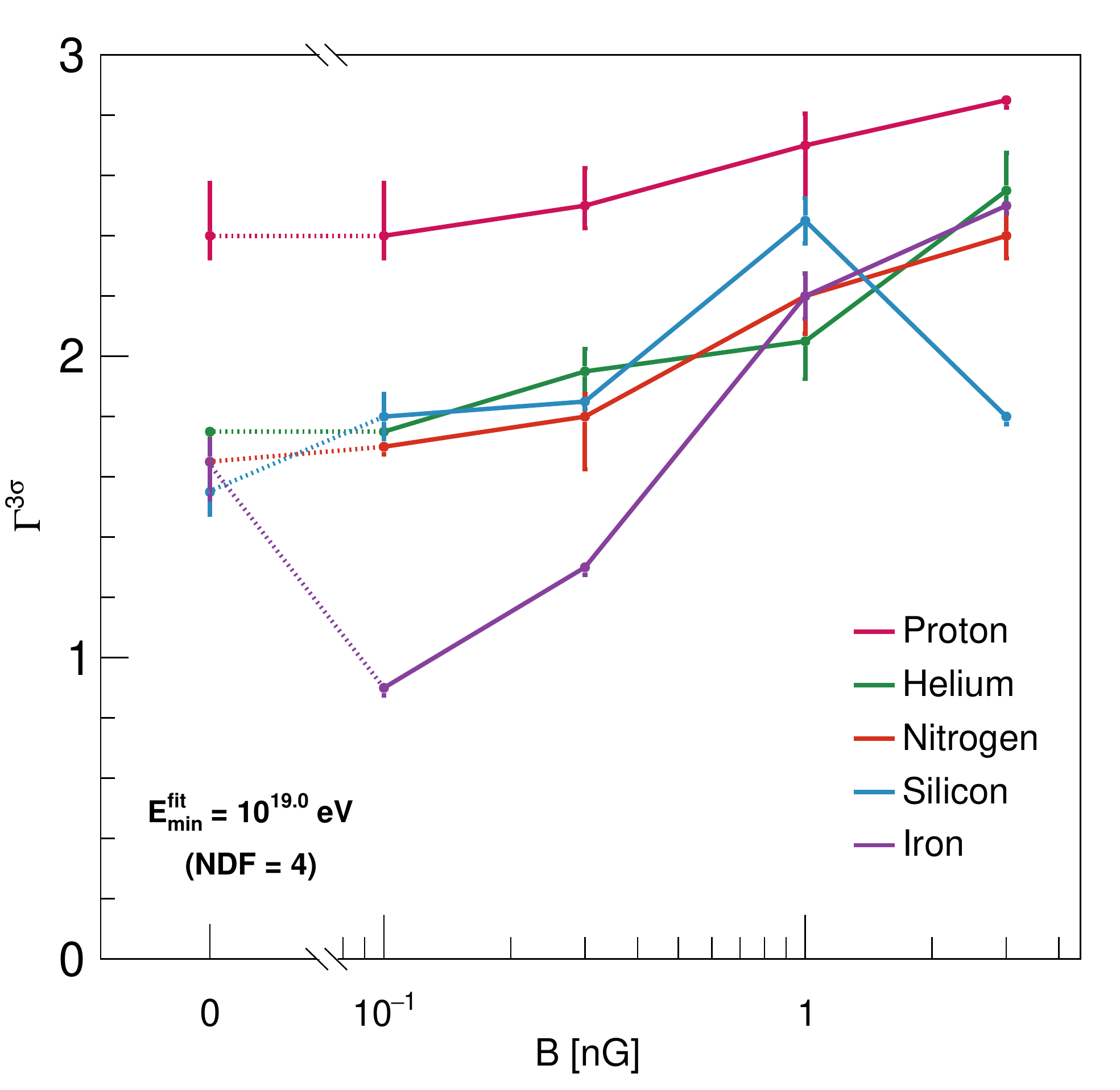}
    \caption{The top panel shows the evolution of the restrained distance to nearest source, $\Dmin^{3\sigma}$ with relation to the field strength, $B$, while the bottom one shows the evolution of the fitted spectral index, $\Gamma$, with relation to $B$. The pink, green, orange, cyan and purple lines represent, respectively, the scenarios with pure proton, helium, nitrogen, silicon and iron composition at the sources.}
    \label{fig:chi2magfields}
\end{figure}

In order to verify the robustness of the results presented in section~\ref{sec:Dmin}, we repeat the fit and constrain the distance to the nearest source under different extra-galactic magnetic field assumptions using the method described in section~\ref{sec:method}. We considered magnetic fields below the upper limits of 3 nG set by observations~\cite{Kronberg_1976,Kronberg:1993vk,Blasi_1999,Schleicher:2011jj}. Additionally, magnetic fields weaker than 0.1~nG have almost no effect within the energy range considered here. Changing $\lcoh$ also impacts the final spectrum. The magnitude of the effect scales with $B\sqrt{\lcoh}$.

Figure~\ref{fig:chi2magfields} shows the evolution of the maximum value of the distance to the nearest source at $3\sigma$ confidence level, $\Dmin^{3\sigma}$, and of the spectral index, $\Gamma$, with relation to the field strength, $B$. In scenarios for extra-galactic magnetic fields within the range allowed by observation that we consider, our results for the least constrained primaries, i.e., nitrogen and iron are strengthened and the results obtained for proton, helium and silicon are confirmed.

Important insight can also be drawn from the evolution of the spectral index, $\Gamma$ with relation to the field strength, $B$. For reasonable magnetic field strengths, a softening of the fitted spectral index is found. This is expected in order to compensate for the hardening of the spectrum coming from the magnetic effects described in section~\ref{sec:horizon}.

From this analysis, as appreciated from figure~\ref{fig:chi2magfields}, we conclude that the existence of sources at $D < 25-100$~Mpc ($z \lesssim 0.02$) are imperative to explain the high-energy end of the spectrum measured by the Pierre Auger Observatory. This corroborates the previous results regarding the need of local sources~\cite{Taylor11}.

\section{Conclusions}

In this work, we have revisited and updated previous studies on the need of local UHECR sources. We have furthered these results by considering the combined effects to the arriving spectrum of magnetic fields and the distance to the nearest UHECR source.

We updated the result of Ref.~\cite{Taylor11} using new data from the Pierre Auger Observatory and adopting the publicly available Monte Carlo propagation code CRPropa 3. A simple fit to the spectral data of the Pierre Auger Observatory for scenarios considering a pure composition at the sources was performed. The fit becomes inconsistent with the data for large distances to the nearest source, $\Dmin$. The resulting upper limits at the 3$\sigma$ confidence level, $\Dmin^{3\sigma}$, as well as the corresponding fit parameters are shown in table~\ref{tab:dmin}. While the fit for proton, helium, nitrogen and silicon favor the scenario of GZK suppression over the scenario of maximum acceleration power of the sources, i.e., large maximum rigidity and soft spectral instead of low maximum rigidity and hard spectral index, the simulations for iron are well described by both scenarios, but slightly favor the latter.

The stability of the analysis was also addressed for the first time. The fit result was shown to be stable with respect to the initial energy bin of the analysis, the primary composition, and the models adopted for the EBL and source evolution. This contrasts with the spectral parameters, which tend to depend strongly on such hypotheses.

For the first time, we studied the combined effects of a distance to the closest source and the presence of turbulent extra-galactic magnetic fields. A semi-analytical method for the propagation of UHECR in these turbulent fields was presented. A mapping of a 1-D simulation into a 4-D simulation using the distance distribution of cosmic rays emitted by a source is used. This method is considerably faster and computationally easier than a full 4-D Monte Carlo simulation.

Although turbulent magnetic fields do not change the total flux from a homogeneous distribution of sources, we show how the contribution of each distance shell is affected. Specifically, we highlight four regimes
for such alteration effects: ballistic, non-resonant scattering enhancement, resonant scattering enhancement, and a low-energy magnetic horizon. Consequently, introducing a reasonable value of $\Dmin$ results in a low-energy suppression and a hardening in the measured spectrum. We discussed the dependence of the horizon on the primary species and the field properties, a change in the composition measurements for lower energies is also expected.

Finally, we have re-analyzed the data accounting for the effects of magnetic fields. For reasonable values of the field strength, our constraints on the distance to the nearest source are confirmed, or even strengthened depending on the primary. Therefore, we reaffirm the previous results that sources at $D < 25-100$~Mpc are imperative to describe the experimental data from the Pierre Auger Observatory. This is consistent with several astrophysical models that predict that the main bulk of the cosmic ray spectrum can be explained by nearby sources~\cite{wibig2007heavy,piran2010new,Biermann_2012,Mollerach_2019}.

In summary, the interplay between magnetic fields and the distance to the nearest source imprints significant features in the spectrum of UHECR. Complimentary to the arrival directions, which contain information about the angular distribution of UHECR sources, the composition and most importantly the spectral data allows to study their radial distribution.

\label{sec:conclusions}

\section*{Acknowledgements}
RGL and VdS acknowledge FAPESP support No.
2015/15897-1, No. 2016/24943-0 and No. 2019/01653-4. MA acknowledges support by \textsc{Villum Fonden} under project no.~18994. RGL and VdS acknowledge the National Laboratory for Scientific Computing (LNCC/MCTI, Brazil) for providing HPC resources of the SDumont supercomputer, which have contributed to the research results reported within this paper. (\url{http://sdumont.lncc.br}). RGL thanks DESY Zeuthen for all the help and infra-structure provided while visiting the institution.

\begin{appendix}

\section{Distance distributions}
\label{app:distributions}

If we consider isotropic random EGMF, the spatial distribution of cosmic rays emitted by a source is only a function of distance, $r$, the diffusive scattering length, $\lambda_{\rm scatt}$ (which depends on the cosmic ray ridigity), and propagation time, $t$. In the following we discuss the probability distribution to observe a cosmic ray test particle at a radial distance $r$ from its source after the propagation time $t$.

Three different regimes are considered. For early times, ($\alpha = 3ct/\lsca < 0.1$), the effects of magnetic fields are still negligible and the propagation is approximately ballistic. A delta distribution is used to describe this regime:

\begin{equation}
\left(\frac{dN}{dr}\right)_{\mathrm{ballistic}} = \delta(r-ct)\,.
\end{equation}
On the other hand, for long travel times, $\alpha > 10$ cosmic ray propagation is well described as a diffusive process. To avoid superluminal propagation, $r > ct$, we use the following truncated Gaussian distribution
\begin{equation}
    \left(\frac{dN}{dr}\right)_{\mathrm{diff}} = \begin{cases}
    A r^2 e^{-\frac{r^2}{2 \sigma^2}}&\mathrm{for} \  r \le ct \\
    0 &\mathrm{for} \ r > ct\
    \end{cases},
\end{equation}
where $\sigma = \sqrt{\lsca ct /3}$ and $A$ is the normalization constant given by

\begin{equation}
    \frac{1}{A}  = \sigma^2 \left(\sqrt{\frac{\pi}{2}} \sigma \mathrm{erf} \left(\frac{ct}{2\sigma}\right) - ct e^{-\frac{(ct)^2}{2 \sigma^2}}\right)\,.
\end{equation}
This distribution describes very well the diffusive regime and is relatively well behaved and easy to treat both numerically and analytically. However, its limit in the ballistic regime ($\sigma \rightarrow \infty$), is given by
\begin{equation}
    \lim_{\sigma\to\infty} \left(\frac{dN}{dr}\right)_{\mathrm{diff}} = 
    \begin{cases}
    \frac{3r^2}{(ct)^3} &\mathrm{for} \  r \le ct \\
    0 &\mathrm{for} \ r > ct
    \end{cases},
\end{equation}
which is not a delta distribution as expected (even though it still peaks at $r = ct$). Therefore, the truncated Gaussian distribution does not describe very well the transition between the ballistic and the diffusive regime.
Consequently, for the transition regime, i.e., $0.1 < \alpha < 10$, a more complex function is needed and the J\"uttner distribution~\cite{Juettner_1911,Aloisio_2009} is used,
\begin{equation}
    \left(\frac{dN}{dr}\right)_{\text{J\"uttner}} 
    = \begin{cases} \frac{r^2 \alpha e^{-\alpha / \sqrt{1 - \left(\frac{r}{ct}\right)^2}}}{(ct)^3 K_{1} \left(\alpha\right) \left(1 - \left(\frac{r}{ct}\right)^2\right)^2}& \mathrm{for} \ r \le ct\\\
    0 &  \mathrm{for} \ r > ct\
    \end{cases},
\end{equation}
where $\alpha = 3 ct / \lsca$. The limits of this distribution for small and large $\alpha$ agree with the ballistic and diffusive regime respectively. Nevertheless, the J\"uttner distribution is much more complex to handle both numerically and analytically.

\begin{figure}[ht!]
    \centering
    \includegraphics[width=0.49\textwidth]{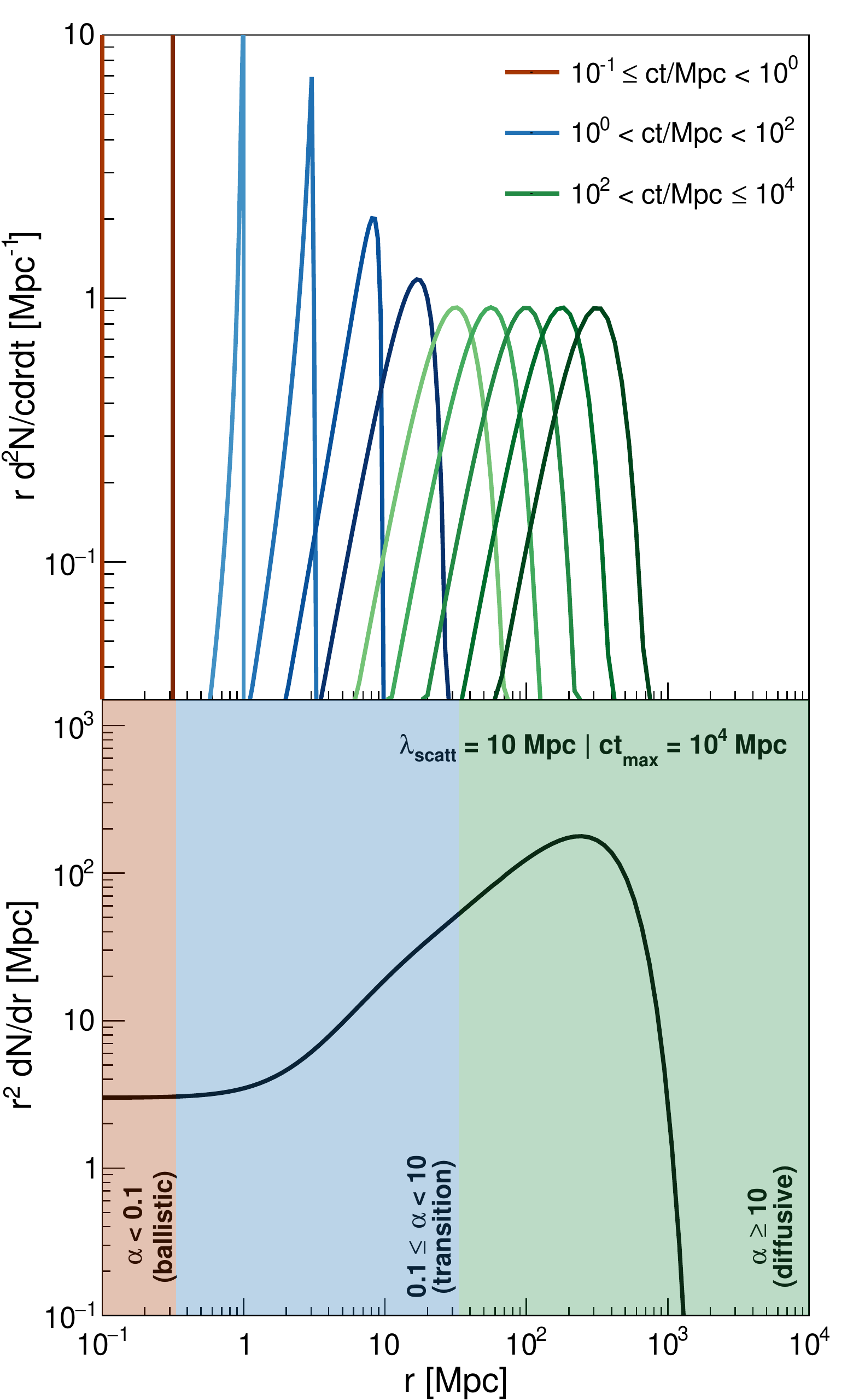}\hfill
    \caption{Distance distribution of cosmic rays emitted by a single source in an environment with turbulent magnetic fields. The top panel shows the time evolution, each line represents the distribution after a given time with log steps. The bottom panel shows the time integrated distribution. In both panels, red, blue and green represent, respectively, the ballistic ($\alpha < 0.1$), transition ($0.1 \leq \alpha < 10$) and diffusive ($\alpha > 10$) regimes. A scattering length, $\lsca=10$~Mpc was chosen and the age of the Universe was taken as $ct_{\mathrm{max}} = 10^{4}$~Mpc.}
    \label{fig:DistanceDistribution}
\end{figure}

Figure~\ref{fig:DistanceDistribution} shows the time evolution of the cosmic ray spatial distribution as well as its integral over time (lower panel). Each of the regimes, in which a different distribution is considered, is highlighted by a different color. For short distances the flux behaves as $1/r^2$, for merely geometric reasons. Farther on, on the diffusive regime, the flux behaves as $1/r$, which is due to the accumulation of events over time. Finally, there is a suppression of the flux due to the finite age of the Universe (in this example plot taken as $ct_{\mathrm{max}} = 10^{4}$~Mpc).

\section{Simulation and Validation}
\label{app:setup}

The simulations were performed with the most widely used package for Monte Carlo simulations of UHECR in the literature, CRPropa 3~\cite{CRPropa3}\footnote{\url{https://crpropa.desy.de}}. The setup consists of a 1-D simulation with no magnetic fields for sources with emitting energy $E_s = [1, \ 10^4]$~EeV and age $ct_s = [1, \ 3162.2]$~Mpc each with 20 bins per decade in $\log_{10}$\footnote{In a 1-D simulation, the age (and consequently travel time) $ct_s = D_s$, where $D_s$ is the distance of the source since all the propagation is ballistic.}. $^{1}$H, $^{4}$He, $^{14}$N, $^{28}$Si and $^{56}$Fe were used as primaries. Pion production, $e^-e^+$ pair production, photodisintegration and adiabatic losses are considered.

For each combination of initial parameters ($E_s$, $t_s$ and primary charge $Z_s$), a number of detected cosmic rays as a function of the energy, $S (E, (E_s, t_s, Z_s))$, was obtained in the simulations. The final flux for each primary and a shell of sources with $D = (D_{\mathrm{min}}, \ D_{   \mathrm{max}})$ is then given with an arbitrary normalization by:
\begin{multline}
    \frac{dN}{dE} (D, E, Z_s) = \sum_{E_s, t_s} S(E, (E_s, t_s, Z_s))  W_{\mathrm{spec}} (E_s, Z_s) \\ 
   \times W_{\mathrm{redshift}} (t_s) W_{\mathrm{mag}} (E_s, t_s, D) W_{\mathrm{sim}} (E_s, t_s)\,,
\end{multline}

where $W_{\mathrm{spec}}$, $W_{\mathrm{redshift}}$, $W_{\mathrm{mag}}$ and $W_{\mathrm{sim}}$ are, respectively, the weights accounting for  spectral features, redshift distribution, magnetic fields and the simulation binning. We use the following ansatz for the (relative) weights
\begin{align}
    W_{\mathrm{spec}} (E_s, Z_s) &\propto E_s^{-\Gamma} e^{E_s/(Z_s\Rmax)}\,, \\
    W_{\mathrm{redshift}} (t_s) &\propto \left(1+z(t_s)\right)^m\,, \\
    W_{\mathrm{sim}} (E_s, t_s)&\propto E_s t_s\,,
\end{align}
where the spectral index, $\Gamma$, and the maximum rigidity at the sources, $\Rmax$ are the spectral parameters. The parameter $m$ accounts for the evolution of the source distribution with redshift and the extra term $E_s t_s$ is needed to compensate for the log binning of the simulation.

The effects of the turbulent magnetic fields are introduced by
\begin{equation}
    W_{\mathrm{mag}} \propto \int_{D_{\mathrm{min}}}^{D_{\mathrm{max}}} \frac{dN}{dr} (\lsca, t_s) dr,
\end{equation}

where $dN/dr$ is the distribution given in Appendix~\ref{app:distributions} and $\lsca$ is given in equation~\ref{eq:lint}.

Finally, the overall CR flux is obtained by the sum shells and primaries:

\begin{equation}
    \frac{dN}{dE} (E) = \sum_{D,Z_s} f(Z_s) \frac{dN}{dE} (D, E, Z_s),
\end{equation}

where $f(Z_s)$ is the fraction of the primary at the source.

We validate our method by a comparison to the results of a full 4-D simulation using CRPropa 3 for a simple case scenario. A pure iron composition at the sources and a strong magnetic field ($B = 2.4$~nG, $\lcoh = 1$~Mpc) are chosen to enhance the effects of the magnetic fields. No energy losses are taken into account in order to avoid masking the effects from the magnetic fields and speed up the simulations. A scenario with $\Gamma = 2$, $\Rmax = 10^{21}$~V and $ct_{\mathrm{max}} = 10
^{4}$~Mpc is considered.

\begin{figure}[h!]
    \centering
    \includegraphics[width=0.49\textwidth]{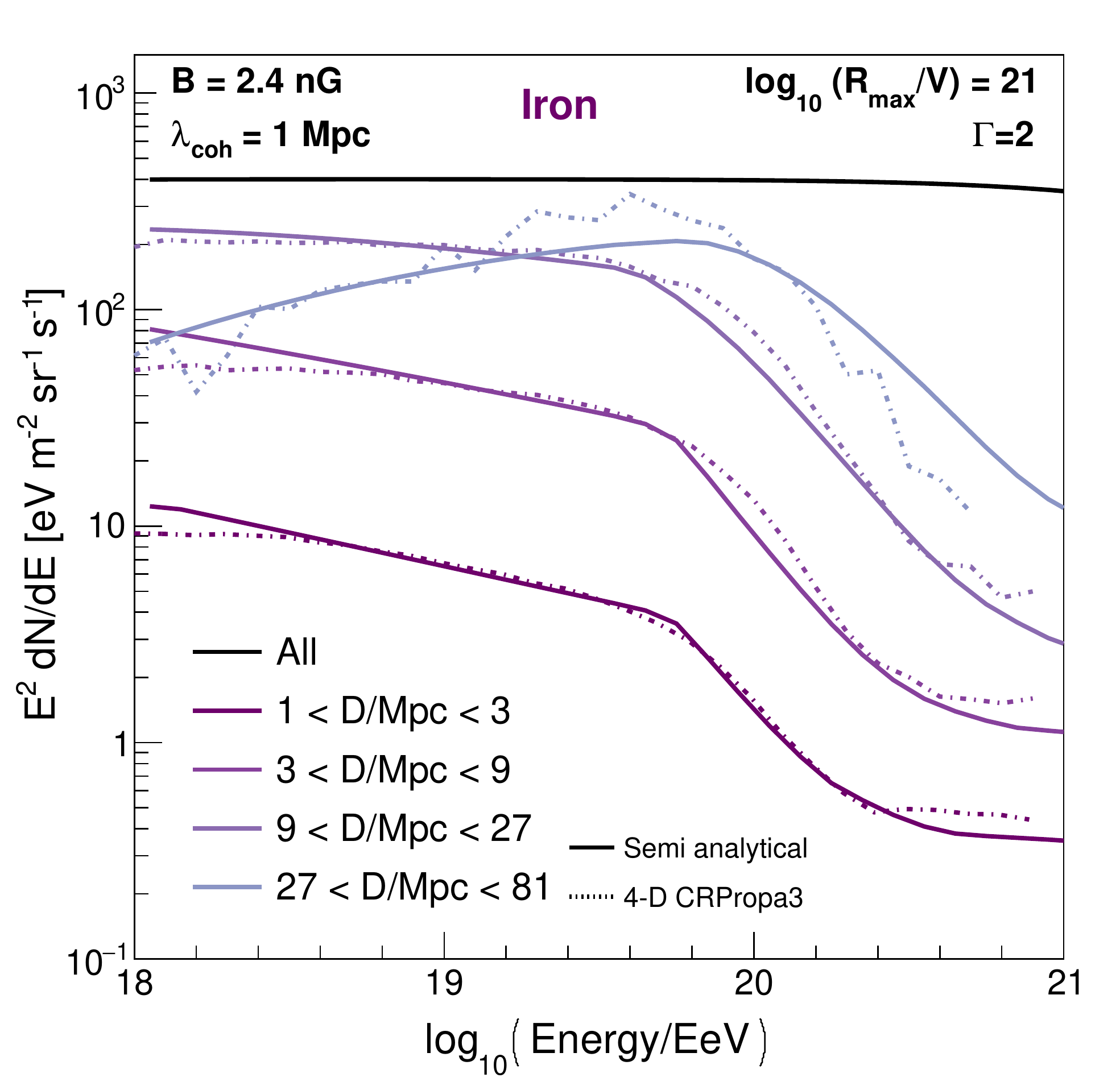}
    \caption{Comparison of the results of the proposed method and a full 4-D simulation. The black line shows the total spectrum while the different colored lines show the contribution of each distance shell. The continuous and dashed lines represent, respectively, the semi-analytical and the full 4-D simulation.}
    \label{fig:crosscheck}
\end{figure}

Figure~\ref{fig:crosscheck} shows the contribution from each distance shell to the total flux obtained with each method. The results are consistent with each other. The results from the 4-D simulation fluctuate much more due to the low statistics of simulated events for farther sources, even with ~$10^4$ times longer simulations. An easier handling of the magnetic field properties is possible with the semi-analytical approach, since this information is contained on the mapping. For the 4-D simulation, on the other hand, it would be necessary to rerun the whole simulation for different field parameters.

Therefore, the proposed method obtains consistent results with a considerably lower computational cost and a better control of the simulation parameters with relation to a full 4-D simulation.

\end{appendix}

\bibliography{references.bib}

\end{document}